\documentclass[letter]{aa}  

\usepackage{graphicx}
\usepackage{txfonts}
\usepackage{lipsum}
\usepackage{subcaption}     
\usepackage{lscape}             
\usepackage{placeins}          
\usepackage[hidelinks=true,hyperfootnotes=true]{hyperref}

\begin{document}

\title{The dispersion in pulsar $\gamma$-ray efficiency}
\authorrunning{Daniel \'I\~niguez-Pascual,
Daniele Vigan\`o, Diego F. Torres}

\author{
Daniel \'I\~niguez-Pascual$^{1,2}$\thanks{E-mail: iniguez@ice.csic.es},
Daniele Vigan\`o$^{1,2}$, Diego F. Torres$^{1,2,3}$
}

\institute{$^{1}$Institute of Space Sciences (ICE, CSIC), Campus UAB, Carrer de Can Magrans s/n, 08193 Barcelona, Spain\\
$^{2}$Institut d’Estudis Espacials de Catalunya (IEEC), 08034 Barcelona, Spain\\
$^{3}$Institució Catalana de Recerca i Estudis Avançats (ICREA), E-08010 Barcelona, Spain }

\abstract
{The observational efficiency of pulsars, defined as the ratio of the observationally derived isotropic-equivalent luminosity, $4\pi d_{obs}^2 F_{obs}$, where $F_{obs}$ is the average pulsed energy flux of a pulsar and $d_{obs}$ is its estimated distance, to its energy budget, shows a wide range of values. This dispersion is believed to be a combination of beaming effects, different geometries, and case-by-case variability of the emission mechanism efficiency, but it is not clear in what proportion.}
{In this work we focused on the $\gamma$-ray range and analysed the four main ingredients that likely contribute to this dispersion: the geometrical term arising from the anisotropic emission (beaming), viewing and inclination angles, the uncertainty on the pulsar distance, the uncertainty on the moment of inertia, and the intrinsic efficiency of the mechanism producing the $\gamma$-ray emission.} 
{Estimating the expected ranges of the moment of inertia and the distance errors, and considering a geometrical and spectral model that we have recently used to fit the light curves and spectra of the entire $\gamma$-ray pulsar population, we estimate the a priori distribution of the first three ingredients in order to obtain the a posteriori distribution of the intrinsic efficiency of the mechanism.}
{We found the latter to peak at $\sim 5-15 \%$ (depending on the trial distribution) and to have a dispersion of around one order of magnitude. That is, we found the intrinsic efficiency of the mechanism to be the leading factor in the observed dispersion. In addition, we found little sensitivity of these results on different distributions of the estimated pulsar distance errors, and saw that the weak, alleged correlation with the spin-down power can only explain part of the observed dispersion. This methodology can be easily applied to other geometrical models of the emission, to test the sensitivity of these results on the beaming distribution.}
{}

\keywords{pulsars: general -- gamma-rays: stars}

\maketitle

\section{Introduction}\label{eta_obs}

The observational efficiency $\eta_{obs}$ is defined as the ratio of the derived isotropic-equivalent luminosity of the pulsar to the pulsar spin-down power:
\begin{equation}
\eta_{obs} = \dfrac{4\pi d_{obs}^2 F_{obs}}{I_{0}\Omega \dot{\Omega}}, 
\label{eq:efficiency_definition}
\end{equation}
where $F_{obs}$ is the average pulsed energy flux of a pulsar, $d_{obs}$ is the estimated distance, $I_{0}=10^{45}$g/cm$^2$ is the conventional value of the moment of inertia, and $\Omega$ and $\Dot{\Omega}$ are the observed spin frequency and its time derivative, respectively. In this work we focus on the $\gamma$-ray range of \emph{Fermi}-LAT (100 MeV -- 300 GeV), for which the values of $\eta_{obs}$ from the Third Fermi Pulsar Catalog (\citealt{3fpc}, hereafter 3PC) show a wide dispersion, with values ranging from $\sim0.02\%$ up to $\sim200\%$. The uncertainty on the pulsar distances and the assumed fixed value of the beaming factor $f_{\Omega} = 1$, are usually mentioned as the origin of this dispersion. Here we consider these two factors, plus the uncertainty on the moment of inertia (neglecting the uncertainties on the timing parameters since they are measured with notable precision), to factorize the combined efficiency as

\begin{equation}
    \eta_{obs} = 
    \frac{4 \pi d_*^2 \langle F \rangle}{I_{*}\Omega \dot{\Omega}}  \frac{I_{*}}{I_{0}} \frac{d^2_{obs}}{d_{*}^2}  \frac{F_{obs}}{\langle F \rangle} \,= \eta_{rad} \,\, \epsilon_I \,\, \epsilon_d \, \, f_{\Omega}^{-1} \,\, , 
    \label{eq:observational_efficiency_extended}
\end{equation}

hereafter conveniently considered in its logarithmic version:

\begin{equation}
    \log_{10}\eta_{obs} = \log_{10}\eta_{rad} + \log_{10}\epsilon_I + \log_{10}\epsilon_d - \log_{10}f_{\Omega} \,\, .
    \label{eq:observational_efficiency_factorized_in_log}
\end{equation}

We label with an asterisk two quantities (the real distance $d_*$ and moment of inertia of the star $I_*$)  which are unknown and might differ from the estimated values, so that
$\epsilon_d := d^2_{obs}/d_*^2$ and $\epsilon_I := I_{*}/I_{0}$ encode the uncertainties on the distance and moment of inertia, respectively. The factor $\eta_{rad} := 4 \pi d_*^2 \langle F \rangle/(I_{*}\Omega \dot{\Omega})$ is the intrinsic efficiency of the mechanism converting the rotational energy losses into $\gamma$-ray emission (via particle acceleration), with $\langle F \rangle$ being the average $\gamma$-ray flux emitted over the sky, i.e. the entire magnetospheric emission in $\gamma$-rays. The beaming factor $f_{\Omega} := \langle F \rangle / F_{obs}$, defined as in e.g. \citet{Watters09}, is the ratio of the total pulsed emission of the pulsar averaged over the sky $\langle F \rangle$ to the phase-averaged emission detected by a given observer $F_{obs}$. This depends on the strongly anisotropic (beamed) emission of the pulsar, the magnetospheric configuration (in particular, the inclination angle $\psi_{\Omega}$ between the rotational axis and the magnetic moment), and the viewing angle $\theta_{obs}$ between the pulsar equatorial plane and the line of sight of the observer. Therefore, the beaming factor can be estimated using a geometrical model of the magnetospheric emission. If the flux intercepted by the observer is exactly the averaged flux over the sky, $f_{\Omega}=1$. Observers in a line of sight that crosses the bulk of the emission have small values $f_{\Omega} < 1$, while those observers that intercept the directions with a lower-than-average flux have $f_{\Omega}>1$. Lucky observers with small beaming factors will estimate a higher observational efficiency than the average observers, which explains the possibility of measuring $\eta_{obs}\gtrsim 100\%$.

In this study we quantified the a priori distribution of values of three terms ($f_{\Omega}$, $\epsilon_d$, and $\epsilon_I$), in order to obtain an a posteriori distribution of the physical mechanism efficiency $\eta_{rad}$, by comparison with the observed distribution of $\eta_{obs}$. We used the standard deviation as a measure of the dispersion of a distribution.

\section{Ingredients contributing to the dispersion of the observational $\gamma$-ray efficiency}

\subsection{Beaming factor, $f_{\Omega}$.}

The beaming factor, $f_{\Omega}$, is here quantitatively estimated using our spectral-geometrical models (for details, see \citealt{compact_formulae,vigano19_light_curves,iniguezpascual24,iniguezpascual25} and references therein). These models were built with an effective approach, have a minimal set of parameters, and are based on following the dynamics of charged magnetospheric particles accelerated by an electric field. The dynamics were computed consistently with the associated synchro-curvature emission. Defining an emitting region that mimics the current sheet shape seen in force-free electrodynamics (FFE) and particle-in-cell (PIC) simulations, we computed the emission maps on the sky, and the light curves for any observer. Synthetic spectra and light curves were fitted to the observed 3PC sample \citep{iniguezpascual24,iniguezpascual25}.

For a given choice of the free spectral parameters and inclination angle $\psi_{\Omega}$, the synthetic sky maps consist of a discrete set of fluxes $F(\theta_{obs,i},\phi_j;\Psi_\Omega)$, where $\phi_j\in[0,2\pi)$ indicates the rotational phase of the pulsar, and $\theta_{obs,i}\in(0,\pi)$ the observer. For this work we used $N_i=51$ observers and a number of phase bins, $N_j$, as defined in \cite{iniguezpascual25}, typically $100$.  We do not discuss here the energy dependence of sky maps and light curves (see \citealt{iniguezpascual25}): fluxes are integrated over the {\em Fermi}-LAT energy range.
The average flux over the sky, $\langle F \rangle$, and the phase-averaged flux detected by the observer, $\theta_i$, $F_{obs,i}$, are

\begin{eqnarray}
    \langle F \rangle &=& \frac{1}{4\pi} \sum_{j=1}^{N_j} d\phi \sum_{i=1}^{N_i} d\theta \, \sin\theta_{obs,i} \, F(\theta_{obs,i},\phi_j;\psi_{\Omega}) \,\, , \\
    F_{obs,i} &=& \frac{1}{2\pi} \sum_{j=1}^{N_j} d\phi \, F(\theta_{obs,i},\phi_j;\psi_{\Omega}) \,\, ,
\end{eqnarray}

and their ratio gives the synthetic beaming factor for each observer, for every sky map. To obtain the $f_{\Omega}$ distribution, we considered the 226 3PC pulsars we fitted in \cite{iniguezpascual25}. For each of them, we generated 30 emission maps, changing $\psi_{\Omega}$, uniformly distributed between $3^{\circ}$ to $90^{\circ}$. For a given sky map, we discarded the observers detecting no flux or a flux below a threshold of   $5\%$ of the maximum of the map, which are about $40\%$ of the total \citep{iniguezpascual25}. Therefore, for every pulsar, we were left with about $\sim 10^3$ synthetic light curves, each one having a value of $f_{\Omega}$. Importantly, the $f_{\Omega}$ distributions obtained from different pulsars are all very similar despite the different spectral and timing ($\Omega$, $\dot{\Omega}$) configurations among the pulsars considered (which implies different free parameters of our spectral model; see \citealt{iniguezpascual25}). This is due to what we found in  \cite{iniguezpascual24}: the geometry ($\psi_{\Omega}$ and $\theta_{obs}$) plays a dominant role in defining the emission maps in $\gamma$-rays, thus effectively decoupling the sky maps (and the $f_\Omega$ distributions) from the observed timing and spectral properties. Therefore, the impact of the geometrical configuration of the pulsar and its emission region on $f_{\Omega}$ overshadows that of the timing and spectral parameters: the $f_\Omega$ distribution as independent of them.

The top left panel of Fig. \ref{fig:histograms_uncertainty_geometrical_terms} shows the histogram of the distribution of the logarithm of $f_{\Omega}$. Not surprisingly, it peaks around $f_\Omega \sim 1$, has a standard deviation of $0.29$ and is asymmetric, with a more pronounced tail at small values rather than large ones (not only due to the logarithmic representation). Although our results might depend on the assumed region emission shape, our $f_\Omega$ distribution looks qualitatively similar to results obtained in other theoretical studies considering either gap models \citep{Pierbattista12} or complex numerical simulations \citep{Kalapotharakos23,Cerutti25}.

\subsection{Uncertainty on the moment of inertia, $\epsilon_I$.}

A conventional value of $I_{0} = 10^{45}$g/cm$^2$ is usually assumed when computing the rotational energy losses. However, each pulsar has its own $I_*$ since it depends on the mass and radius, although these are generally unknown. In addition, the equation of state (EoS) for cold dense matter is still not known precisely, which is reflected in the uncertainty on the neutron star (NS) characteristic mass-radius ($M(R)$) relation.

To take into account both effects, we can estimate $I_{*}$ using the expression presented by \cite{Bejger02}, which, considering a large sample of NS structures built with many different available EoSs and masses, found a best-fitting effective expression for the moment of inertia: $I = a(x) M R^2$, with $x = (\text{M/M}_\odot) (\text{km/R})$ and $a(x) = x/(0.1 + 2x)$ for $x \leq 0.1$ and $a(x) = \frac{2}{9}(1 + 5x)$ for $x > 0.1$. In order to obtain an informed a priori distribution of values of $I_{*}$, we also need the distribution of $M$ and $R$ of NSs. Masses can be estimated in several cases, mainly when in binary systems \citep{Ozel16}, and their values are well described by a skew normal distribution, $\mathcal{SN}$ \citep{Kiziltan13}, whose parameters are the mean $\mu$, the standard deviation $\sigma$, and the skewness $\alpha$.  Using the best-fitting parameters to the most updated available sample,\footnote{\url{www3.mpifr-bonn.mpg.de/staff/pfreire/NS_masses.html}} we then assume an a priori distribution of the masses (in solar masses) $P(M[M_\odot]) = \mathcal{SN}(\mu=1.20, \sigma=0.31, \alpha=14.03)$. Given the substantial lack of precise constraints on the radius, we take $R$ to be distributed as a Gaussian $P(R{\rm [km]})=\mathcal{N}(\mu=12.0, \sigma=1.0)$,  a representative distribution, where the dispersion assumed here is compatible with the relatively vertical shape of the $M(R)$ curve for most EoSs. Given the lesser role played by the moment of inertia (see below), we do not explore other mean values of $R$.

\begin{figure}
        \centering
        \includegraphics[width=\columnwidth]{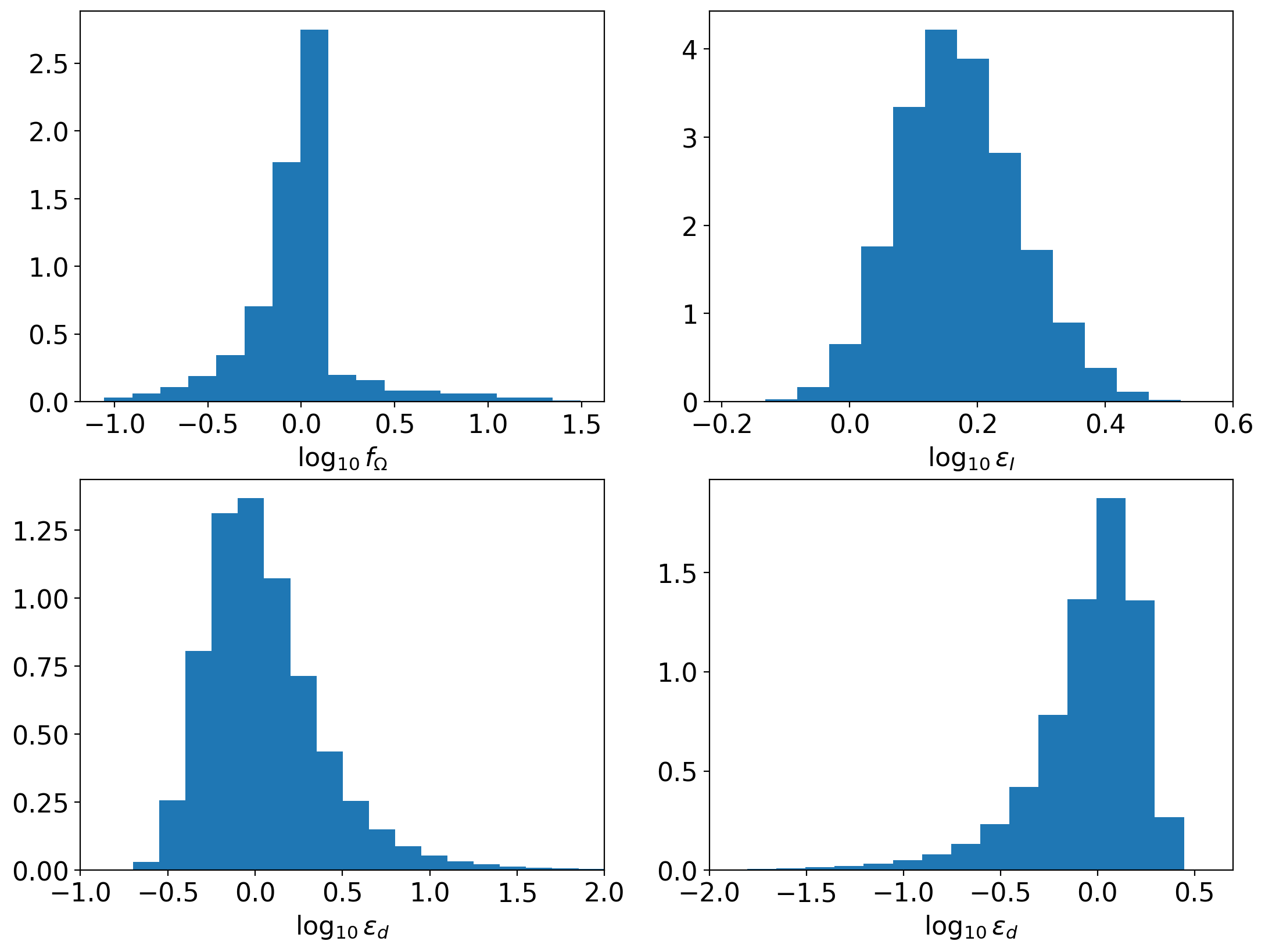}
    \caption{Probability density of the terms forming $\log_{10}\eta_{obs}$. Top: $\log_{10}f_{\Omega}$ (left) and $\log_{10}\epsilon_I$ (right). Bottom: $\log_{10}\epsilon_d$ from different prescriptions, $P(\Delta)=\mathcal{N}$ (left) and $P(\epsilon_d)=\mathcal{N}$ (right).}
    \label{fig:histograms_uncertainty_geometrical_terms}
\end{figure}

Sampling ($\sim$10$^3$) values of $R$ and $M$ from the mentioned distributions with an acceptance-rejection method, we can obtain a distribution of physically motivated values of the moment of inertia. Dividing it by $I_{0}$ gives the uncertainty term associated with the moment of inertia $\epsilon_I$. The right top panel of Fig. \ref{fig:histograms_uncertainty_geometrical_terms} shows the histogram of the distribution of its logarithm. It is centred at $\log_{10}\epsilon_I\sim0.2$ and has a standard deviation of $0.09$, smaller than that of the distributions $\log_{10} f_{\Omega}$ and $\log_{10} \epsilon_d$ (see below).

\subsection{Uncertainty on the distance, $\epsilon_d$}\label{uncertainty_distance}

The real distance to a pulsar from Earth, $d_{*}$, is unknown.  The observationally estimated values, $d_{obs}$, normally have large uncertainties and are obtained with different methods, for example  the dispersion measurement method, parallax, or kinetic method.  As a proxy of the distance uncertainties distribution, we take the average of the relative distance errors of the pulsar sample, as quoted in the 3PC, which is $\sigma_{\rm d,rel}=0.341$.  In the absence of a better estimation for the error distribution, we consider two cases.  In the first case, we assume that the relative error  $\Delta := (d_{*} - d_{obs})/d_{obs}$ follows a Gaussian distribution, with $\mu=0$ and $\sigma=\sigma_{\rm d,rel}$. We then sample ($\sim 10^6$) $\Delta$ values with an inverse transform method, and obtain the corresponding set of values $\epsilon_d=(\Delta+1)^{-2}$. The second choice is to instead consider the distribution of $\epsilon_d$ itself as a Gaussian (sampling $\sim 10^6$ values), with $\mu=1$ and $\sigma_{\epsilon,d}=(\sigma_{d,rel}+1)^{-2}=0.556$. The resulting $P(\log_{10}\epsilon_d)$ for the two choices (left panels of Fig. \ref{fig:results_fitting_mechanism}) differ in shape, but only slightly in standard deviation ($0.38$ and $0.31$, respectively).

\section{Distribution of the mechanism efficiency}\label{efficiency_mechanism}

The rotational energy losses represent the budget powering pulsar high-energy emission. However, which fraction is ultimately converted into emitted radiation is not known exactly since it depends on the details of every step of the conversion: Poynting flux $\rightarrow$ particle creation and acceleration $\rightarrow$ synchro-curvature radiation (in $\gamma$-ray range).

We assumed a trial distribution for $\log_{10}\eta_{rad}$ and, assuming that the four distributions on the right-hand side of Eq. \eqref{eq:observational_efficiency_factorized_in_log} are independent of each other, their sum (hereafter referred to as joint distribution) is simply obtained by performing a Monte Carlo sampling of each distribution and summing the values (see e.g. Chapter 3 of \citealt{algebra_random_variables}).  We then compared it with the distribution of the logarithm of the observational efficiency $\log_{10}\eta_{obs}$ by applying a Kolmogorov--Smirnov (KS) test. The metric of the KS test, $D_{KS}$, measures the similarity between two distributions, by using their empirical distribution functions. We minimized $D_{KS}$ by scanning the space of parameters of the trial distribution and picking up the set of parameters giving a smaller $D_{KS}$. The best-fitting set of parameters defining the distribution of $\log_{10}\eta_{rad}$ is the one that generates a joint distribution most closely  resembling $\log_{10}\eta_{obs}$. Finally, we obtained the dispersion and peak position of the best-fitting trial distribution.

\begin{figure*}
        \centering
        \includegraphics[width=\textwidth]{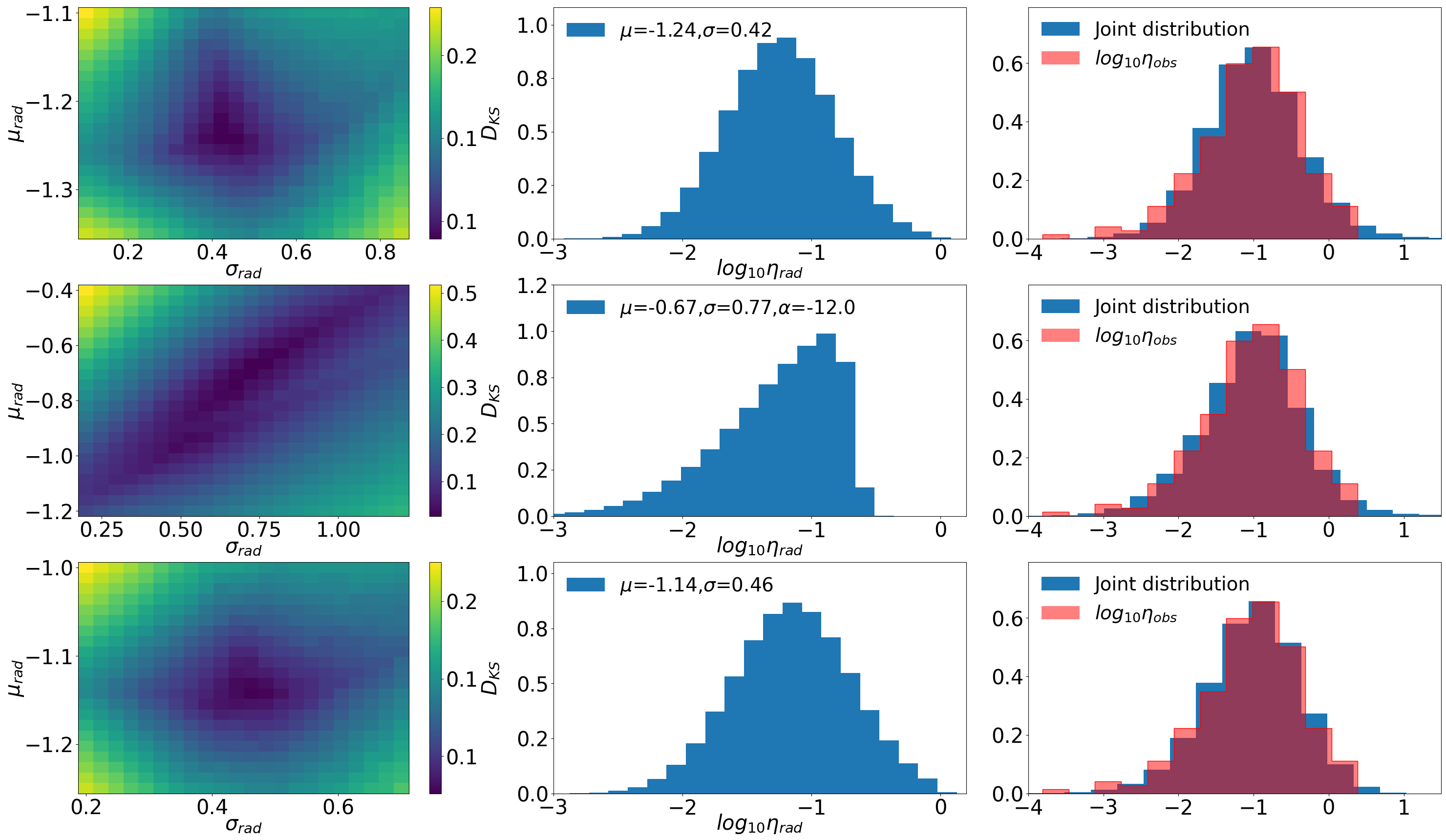}\\
    \caption{Results of the fitting to find the distribution of $\eta_{rad}$. The top (middle) row shows those obtained using the distance uncertainty distributed as $P(\Delta)$=$\mathcal{N}$, with a Gaussian (skew normal) as the trial distribution. The bottom row were obtained with the distance uncertainty distributed as $P(\epsilon_d)$=$\mathcal{N}$, with a Gaussian as the trial distribution. From left to right: contour plot of $D_{KS}$ in the space of parameters $\mu_{rad}$-$\sigma_{rad}$ (in the skew normal case we collapsed all the contours for the different $\alpha$ considered taking the smallest $D_{KS}$ for each pair $\mu_{rad}$-$\sigma_{rad}$), best-fitting distribution of $\log_{10}\eta_{rad}$, overlapped distribution of $\log_{10}\eta_{obs}$ (red), and joint distribution (blue).}
    \label{fig:results_fitting_mechanism}
\end{figure*}

Table \ref{tab:best_fitting_eta_rad} compares the best-fitting parameters of the trial distributions (Gaussian or skew normal) of $\log_{10}\eta_{rad}$ for the two different prescriptions of distance uncertainties. The resulting joint distributions are shown in the right column of Fig. \ref{fig:results_fitting_mechanism}, compared with $\log_{10}\eta_{obs}$. In all cases, the dispersion of $\log_{10}\eta_{rad}$ is similar, with standard deviations of $0.42-0.56$, which corresponds to around one order of magnitude in $\eta_{rad}$, regardless of the different shapes of distributions. The best-fitting a posteriori distributions peak at a value $\eta_{rad}^{peak}$, which changes between $\sim 5$ and $\sim 15\%$;   the change is mainly due to the different shapes of the trial distributions.

Note that in all cases there is a slight mismatch at low $\eta_{obs}$ values. This can be produced by the observational biases that affect this region: the small values may be a result of a low flux close to the $\emph{Fermi}$-LAT sensitivity. Thus, the comparison at low values is not informative and we do not focus on them.  Instead, note that the cases with $P(\Delta)= \mathcal{N}$ have a tail at high values ($\log_{10}\eta_{rad}>10$), which are incompatible with the observed distribution regardless of the shape of the distribution $P(\log_{10}\eta_{rad})$, and is due to cases with very large $\log_{10}\epsilon_d \gtrsim 1$ (i.e. an order of magnitude overestimation of $d_*^2$).  Such a tail is indeed much less visible if we consider the distance uncertainty distributed as $P(\epsilon_d) = \mathcal{N}(1,\sigma_{\epsilon,d})$ because the latter has fewer extremely high values. Moreover, for the latter case, we   explored the sensitivity of the fitting on the dispersion of the a priori distribution $P(\epsilon_d)$ by taking $\sigma_{\epsilon,d}=0.2$ and $0.7$ instead of $0.556$. The result is that the peak position of the mechanism efficiency distribution barely changes, while its standard deviation change by only  $\sim 30\%$ between the two extremes of $\sigma_{\epsilon,d}$.

Overall, this indicates a minor influence of the $\epsilon_d$ prescription on the best-fitting a posteriori distribution of $\eta_{rad}$. It also implies that $\eta_{rad}$ is the term dominating the dispersion of $\eta_{obs}$.

\begin{table}[t]
    \centering
    \caption{Best-fitting parameters of the trial distributions of $P(\log_{10}\eta_{rad})$ for different distributions of distance uncertainty.}
    \begin{tabular}{cccc}
        $P(\log_{10}\eta_{rad})$ & Distance uncertainty & $\eta_{rad}^{peak}$ & Std dev \\\hline\hline
         ${\cal N}$ & $P(\Delta)={\cal N}(0,0.341)$ & $5.8\%$ & $0.42$ \\
         ${\cal SN}$ & $P(\Delta)={\cal N}(0,0.341)$ & $14.8\%$ & $0.47$ \\
         ${\cal N}$ & $P(\epsilon_d)={\cal N}(1,0.556)$ & $7.2\%$ & $0.46$ \\
         ${\cal N}$ & $P(\epsilon_d)={\cal N}(1,0.2)$ & $6.6\%$ & $0.56$ \\
         ${\cal N}$ & $P(\epsilon_d)={\cal N}(1,0.7)$ & $7.1\%$ & $0.42$ \\
    \end{tabular}
    \label{tab:best_fitting_eta_rad}
\end{table}

\section{Discussion}\label{discussion}

The dispersion of the observational $\gamma$-ray efficiency $\eta_{obs}$ can be explained with the four ingredients we have introduced here: $f_{\Omega}$, $\epsilon_d$, $\epsilon_I$, and $\eta_{rad}$. We found $\eta_{rad}$ to be a main contributor, at least as relevant as the distance errors and the $f_\Omega$ distribution coming from our geometrical model. The contribution of $\epsilon_I$ is smaller compared to the other three terms, which indicates that assuming $I=10^{45}$g/cm$^2$ for all pulsars does not significantly affect the observational efficiency values. With the two trial distributions we   considered for the efficiency of the physical mechanism of $\gamma$-ray emission $\eta_{rad}$, we   found a peak position of around $5-15\%$, with a dispersion of approximately one order of magnitude. The same conclusion was obtained for a few different trial distributions of $\eta_{rad}$ and/or prescriptions regarding the uncertainty on the pulsar distances.

The average values we found agree with the results found in complex magnetospheric simulations (FFE or PIC), in which $\sim 10-20\%$ of the Poynting flux is dissipated in the first $1-2$ light cylinder radius of the equatorial current sheet \citep{Carrasco18,Cerutti20,Hakobyan23}. The dissipated Poynting flux in the current sheet is converted into acceleration of particles, which then emit the high-energy radiation we detect. In our spectral-geometrical models, the $\gamma$-ray emission typically represents $\gtrsim 95\%$ of the total synchro-curvature radiation, in agreement with the tiny (always less than $1\%$) X-ray observational efficiencies of Fermi pulsars reported in e.g. \cite{Marelli11}. The mean values of $\eta_{rad}$ also agree with the fact that pulsar wind nebulae, which are powered by the pulsar itself, require a significant amount of its energy.

The relatively large dispersion of the inferred $\eta_{rad}$ distribution could be due to different factors. The alleged weak correlation $\eta_{obs} \propto (I_0\Omega\dot\Omega)^{-0.5}$ reported in the 3PC can explain a part of the dispersion (see Appendix \ref{correlation_etaobs_edot}). Other sources of dispersion could be differences in e.g. magnetic field topology at the surface or beyond the light cylinder (boundary conditions with the surrounding medium). General relativity effects on the torque and pair production, which depend on the compactness \citep{Ruiz14,Philippov15,Carrasco18}, could also be a minor ingredient for dispersion. Note that the precise shape of the distribution of $\eta_{rad}$ remains unconstrained with our methodology since a good fit is obtained with the two trial distributions. In this regard, note that the tail with very high arguably unphysical values $\gtrsim 50\%$ of efficiency is actually not needed: imposing a skew normal distribution, or, similarly, a cutoff at $\eta_{rad} \gtrsim 30\%$, essentially provide the same results in terms of inferred mean and dispersion.

Finally, this simple methodology can be implemented to test whether different geometrical models, with their own $f_{\Omega}$ distributions, can change the results here presented.

\begin{acknowledgements}
      We thank the referee David Smith for his valuable comments.
      This work has been supported by PID2024-155316NB-I00 funded by MCIN/AEI/10.13039/501100011033, CSIC PIE 202350E189, the Spanish program Unidad de Excelencia María de Maeztu CEX2020-001058-M and  by European Union NextGeneration EU (PRTR-C17.I1). DIP has been supported by the FPI pre-doctoral fellowship PRE2021-100290 from the Spanish MCIU and his work has been carried out within the framework of the doctoral programme in Physics of the UAB. DV is funded by the European Research Council Starting Grant IMAGINE, No. 948582.
\end{acknowledgements}

\bibliographystyle{aa}
\bibliography{dispersion_gamma_ray_efficiency}

\onecolumn
\begin{appendix}
    \section{Additional sources of $\eta_{obs}$ dispersion}\label{correlation_etaobs_edot}

\begin{figure*}[h]
        \centering
        \includegraphics[width=\textwidth]{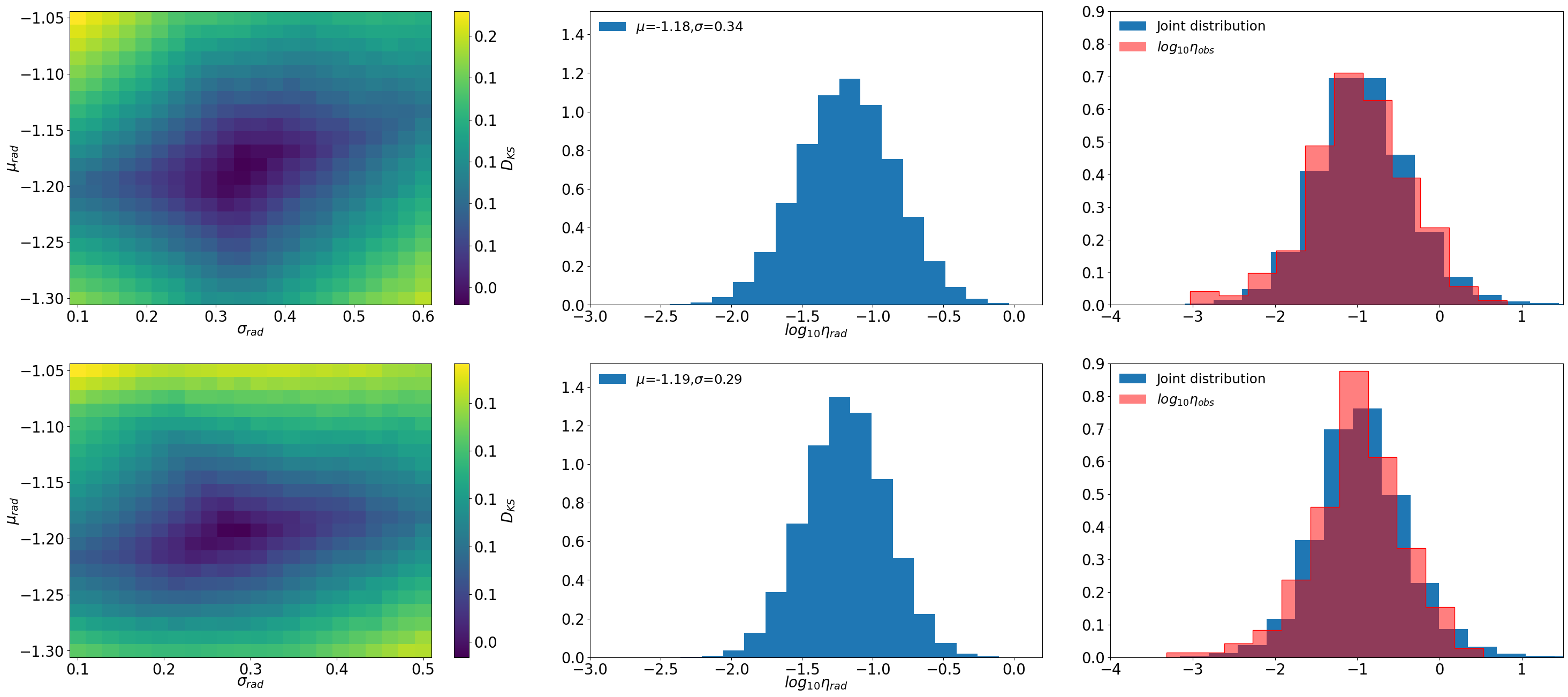}\\
        \caption{Results of the fitting to find the distribution of $\eta_{rad}$, 
        analogous to Fig. \ref{fig:results_fitting_mechanism}. The top (bottom) row shows those obtained with a $\log_{10}\eta_{obs}$ distribution de-trended from the correlation $\propto \dot{E}^\beta$, with $\beta$ equal to $-0.5$ ($-0.32$). Both cases were obtained using the distance uncertainty distributed as $P(\Delta)$=$\mathcal{N}$ and with a Gaussian as the trial distribution.}
        \label{fig:results_fitting_mechanism_detrended}
    \end{figure*}

    The 3PC, similarly to previous studies, proposes a correlation between the spin-down power $\dot{E}=I_0\Omega\dot \Omega$ and the observational efficiency $\eta_{obs}$: $\eta_{obs} \propto \dot{E}^{\beta}$, with $\beta=-0.5$ (see Fig. 24 of the 3PC). This correlation is not strong (the Pearson r coefficient in the log-log relation is $-0.48$), and a much larger sample over a broader range of $\dot{E}$ would be needed for a proper assessment (see e.g. \citealt{Szary14} in the X and radio bands). However, since the weakness of the correlation can be indeed due to the other concurrent dispersion factors here studied, we have investigated the contribution of the alleged correlation to the dispersion in the observational efficiency. By removing the correlation from the observed distribution, the de-trended $\log_{10}\eta_{obs}$ distribution has a slightly smaller dispersion than the original one (standard deviation of $0.60$ and $0.66$, respectively). If we let $\beta$ free instead of fixing it to -0.5, we find a best-fitting relation of $\eta_{obs} \propto \dot{E}^{-0.32}$, which leads to a de-trended distribution of $\log_{10}\eta_{obs}$ with a slightly smaller standard deviation, $0.57$. In order to see how this smaller dispersion affects the inferred $\eta_{rad}$ distribution, we have repeated the procedure presented in Sect. \ref{efficiency_mechanism}. The corresponding best-fitting distribution of the efficiency mechanism $\eta_{rad}$ are shown in Table \ref{tab:best_fitting_eta_rad_detrending} and Fig. \ref{fig:results_fitting_mechanism_detrended} show the fitting results, with a Gaussian as a trial distribution for $\eta_{rad}$ and the $\epsilon_d$ prescription in which $\Delta$ is distributed as a Gaussian. The dispersion of $\eta_{rad}$ slightly decreases, as expected from the smaller dispersion of the de-trended $\eta_{obs}$, while the central value is basically unaffected. These results indicate that a non-negligible, but not dominant part of the inferred $\eta_{rad}$ dispersion can be attributed to the variations in $\dot{E}$. In other words, the $\gamma$-ray luminosity might scale less than linearly with $\dot E$, but this alone cannot explain the bulk of the inferred $\eta_{rad}$ dispersion.

    \begin{table}[h]
        \centering
        \caption{Best-fitting parameters of $\log_{10}\eta_{rad}$ with $P(\log_{10}\eta_{rad}) = {\cal N}$ and $P(\Delta)={\cal N}(0,0.341)$ for the distance uncertainty, considering the original $\log_{10}\eta_{obs}$ distribution or the ones de-trended with the correlation $\propto \dot{E}^\beta$.}
        \begin{tabular}{cccc}
            $\log_{10}\eta_{obs}$ distribution & $\log_{10}\eta_{obs}$ std dev  & $\eta_{rad}^{peak}$ & $\log_{10}\eta_{rad}$ std dev \\\hline\hline
             Original & $0.66$ & $5.8\%$ & $0.42$ \\
             De-trended, $\beta=-0.50$ & $0.60$ & $6.6\%$ & $0.34$ \\
             De-trended, $\beta=-0.32$ & $0.57$ & $6.5\%$ & $0.29$ \\
        \end{tabular}
        \label{tab:best_fitting_eta_rad_detrending}
    \end{table}

    \end{appendix}

\end{document}